\documentclass[a4paper,11pt]{article}
\pdfoutput=1 

\usepackage{jinstpub} 

\usepackage{lineno}
\usepackage{caption}
\usepackage{subcaption}

\title{\boldmath Meeting the challenge of Open Science in KM3NeT}

\author[a]{J. Schnabel \note{Corresponding author.}}
\author[b]{P. Kalaczy\'{n}ski}
\author[c]{C. Bozza}
\author[a]{T. Gal}

\affiliation[a]{ECAP, Friedrich-Alexander-Universit\"at Erlangen-N\"urnberg,\\Erwin-Rommel Str. 1, 91058 Erlangen, Germany}
\affiliation[b]{NCBJ - National Centre for Nuclear Research,\\ul. Pasteura 7, Warsaw 02-093, Poland}
\affiliation[c]{Universit\`{a} di Salerno e INFN,\\Via Giovanni Paolo II 132, Fisciano 84084, Italy}

\emailAdd{jutta.schnabel@fau.de}

\abstract{In the upcoming decades, the KM3NeT detectors will produce valuable data that can be used in various scientific contexts from astro- and particle physics to environmental and Earth and Sea science. Based on the Open Science policy established by the KM3NeT Collaboration, several efforts to offer science-ready data, foster common analysis approaches and publish open source software are currently pursued. In this contribution, ongoing projects focusing on the exchange of high-level data and simulation derivatives, production of particle event simulations and establishment of an integrated computing environment supporting an open-science focused workflow will be discussed.}


\collaboration[c]{on behalf of the KM3NeT collaboration}

\proceeding{VLVnT2021\\
  18-21 May 2021\\
  Valencia}

\begin{document}
\maketitle
\flushbottom

\section{From FAIR data to FAIR research}

\subsection{The FAIR principles}
The FAIR principles \cite{FAIR} to make scientific data Findable, Accessible, Interoperable and Reusable are starting to become a new standard of scientific practice. While these principles are primarily applied to data with the goal to maximize the scientific benefit from available research data, they can also be understood as a requirement for efficient research management already at the stage of primary data selection and science project layout, and in the build-up of common multi-messenger analyses from multiple experiments.

The KM3NeT collaboration has presented the prototype of an open science system \cite{OSS} which enables data sharing with the Virtual Observatory ecosystem\footnote{Provided by the IVOA: \url{https://www.ivoa.net/}.} for astronomical analysis and aims at providing data through public repositories like Zenodo\footnote{Of the OpenAIRE project: \url{https://zenodo.org/}.} also for the particle physics community. 
Beyond data sharing, the system also includes resources for teaching and sharing of software and workflows. This procedural knowledge is key to the usability of the research data, and leads to the necessity to introduce a modern and agile approach to software development and sharing, including automated building and testing of software to ensure software quality. 

The collaborative environment offered by development platforms like Gitlab\footnote{\url{https://about.gitlab.com/}, with KM3NeT offering software and containers via \url{https://git.km3net.de/}, or on github: \url{https://github.com/KM3NeT}.} also has a positive impact on analysis projects as such, as it allows an easy sharing of analysis scripts and setups in the same style as code development projects, and thus enables detailed feedback and a transparent access to all steps of the analysis process. 

While this transparency of the research processes is already beneficial for the internal quality control of a collaboration, it is even more helpful for the development of common research practices between collaborations. Therefore, the KM3NeT collaboration actively develops its open science system with this broadened application of the FAIR principles in several internal and external initiatives focusing on the research process and common development not only of data format standards, but common software, services and an increasingly integrated research environment.

\subsection{Internal standards through FAIRness of research}
In order to interpret and apply FAIR standards within the collaboration, a dedicated Open Science Committee (OSC) has been created aiming at setting and fostering standards on software and data publication. While, classically, quality control focuses on the published paper, scientific reasoning and high-level results, the OSC focuses on the means used in the analysis process and ensures that the process is reproducible and can be checked. This includes the use of verified releases of data and software, and the availability of the analysis scripts for testing and cross-checks.
 
\subsubsection{Data processing}
One of the larger difficulties of data processing is designing a pipeline which runs on a large variety of systems. An analysis usually starts on a scientist's desktop and is gradually extended and improved until the point when the first batch processing can be executed on a larger set of machines. While a single computing centre might offer a uniform architecture and environment, the next evolution of the analysis could involve multiple batch systems. Additionally, the full data processing chain should remain reproducible in future. In order to limit the maintenance efforts and make the processing at least to a good extent system agnostic, the Nextflow\cite{Nextflow} workflow management system is currently being integrated in KM3NeT which only depends on Java and provides a domain-specific language to describe workflows and processes. It supports many popular batch processing systems and allows the execution of each step in containerized environments using either Docker or Singularity, which are used in KM3NeT use to develop, test and deploy software. Therefore, a full data processing chain can be designed using Nextflow and different configuration profiles make it possible to execute them on either a single desktop or a large computing centre with tens of hundreds of nodes.

For a distributed computing environment, not only workflow management, but also a systematic data management is crucial, enabling FAIR data sharing. In this as well as in other initiatives towards open science, KM3NeT participates in the developments of the ESCAPE project\footnote{European Science Cluster of Astronomy and Particle physics ESFRI research infrastructures \url{https://projectescape.eu/}.} and explores also the internal use of the Rucio\footnote{\url{https://rucio.cern.ch/}} scientific data management software.

\subsubsection{Analysis procedures}
To harmonize the analysis process, template Gitlab repositories are provided and developed which make the use of continuous integration, automated code documentation and providing of a containerized version of the repository using Docker\footnote{On the Docker products for virtual containers, see \url{https://www.docker.com/}.} easier for researchers. While these repositories only contain processing scripts for the usually large amount of data available in a high-performance computing infrastructure and used as basis for the analysis, the highest-level data selection used for the calculation or display of the final results in a publication are made available in the repository. This enables cross-checks for members of the collaboration already during the genesis of the analysis.

\section{Initiatives to enhance Open Science in KM3NeT}
In the ESCAPE project, the EOSC\footnote{European Open Science Cloud, \url{https://eosc-portal.eu/}.} use for astro-, astroparticle and particle physics experiments is developed. As this scientific environment includes not only a solution for common data storage and access, but also an open software and service repository and a science analysis platform, and focuses also on the development of Virtual Observatory standards, it offers the ideal environment to enhance the understanding of sharing and set up of open scientific workflows.

\subsection{Open Software}
An example of a code developed within the KM3NeT Collaboration, and made open to the public is gSeaGen \cite{gSeaGen}, a GENIE-based application for neutrino telescopes \cite{GENIE}. It is designed to simulate neutrino flux at the detector using the neutrinos generated by GENIE. 

The gSeaGen source code is publicly available on the Zenodo platform \cite{gSeaGen-ZENODO}. With user comfort in mind, efforts are made to distribute it in the form of docker containers. For the same reason, there is an ongoing discussion among the gSeaGen developers about the possible alternatives for the public output format (currently, only ROOT trees are supported).

In the near future, it is foreseen to add the functionality to propagate the muons (and later also neutrinos) generated with CORSIKA \cite{CORSIKA}.

Open software from KM3NeT is planned for integration in the ESCAPE Open Software and Service Repository, which also will include python-based software organized in the km3py meta package\footnote{Available through PyPI: \url{https://pypi.org/project/km3py/}.}.

\subsection{ConCORDIA}
Populating the EOSC with ready-for-use solutions is an effective way to foster the sharing of knowledge and expertise across diverse communities. A joint effort has started between CTA and KM3NeT to produce a comfortable environment for easy simulation of cosmic rays based on the CORSIKA code. Both collaborations identified a set of use-cases, originating from the needs to effectively represent the instrument response functions of the detectors, and hence exploring several alternative setups, mostly differing for the hadronic interaction model and the production channels of massive secondaries (charms, tauons). Each use-case is hosted in a separate container, which can then be invoked immediately to produce reasonable outputs. Containers will be published on the EOSC with accompanying statistical graphs assessing the data quality and performance. The ConCORDIA environment also includes a Web Graphical User Interface for container customization and job management on DIRAC: CORSIKA running parameters can be easily tuned even by non-experts. The common development to support the usage of the DIRAC interware is aimed at providing a common, unified solution that can be used regardless of the specific computing infrastructure the user can access: DIRAC has the ability to drive resources that are distributed among several data centers, which is optimally suited for the EOSC scenario of researchers coming from different places and different institutions.

\subsection{Developing standards}
The ESCAPE project bases developments on use cases of scientific analyses, especially highlighting cross-experiment initiatives. These combined workflows require a special focus on data format standards, for which especially overlapping features with gamma ray astronomy are explored. The established gamma-ray astronomy standard for ground-based imaging atmospheric Cherenkov telescopes\footnote{Data formats for gamma-ray astronomy, \url{https://gamma-astro-data-formats.readthedocs.io}.}, which also involves instrument response functions, seems to meet the needs for neutrino detectors already to a large extent. In cooperation with the IVOA, a further development of a common high-level data format is aimed for in ESCAPE.\\

The efforts by the KM3NeT collaboration to create an environment for open science are guided by the FAIR principles. However, by applying these principles to the full research process, reusability and transparency can also be increased in the development of software, management of workflows and sharing of work in areas of common research interests. The examples given here showcase potential areas of development to help establish a culture of open science in the community.

\end{document}